\documentclass[journal=jctcce,manuscript=article]{achemso}

\usepackage{graphicx}
\usepackage{color}
\usepackage{epstopdf}
\usepackage{amsthm}
\usepackage[tight]{subfigure}

\newcommand{\bfr}{{\bf r}}

\title{The Particle-Hole Map: a computational tool to visualize electronic excitations}

\author{Yonghui Li}
\affiliation{Department of Physics and Astronomy, University of Missouri, Columbia, Missouri, 65211, USA}
\alsoaffiliation{Department of Physics and Astronomy, Vanderbilt University, Nashville, Tennessee, 37235, USA}
\author{Carsten A. Ullrich}
\email{ullrichc@missouri.edu}
\affiliation{Department of Physics and Astronomy, University of Missouri, Columbia, Missouri, 65211, USA}

\begin{document}

\begin{abstract}
We introduce the particle-hole map (PHM), a visualization tool to analyze electronic excitations in molecules in the time
or frequency domain, to be used in conjunction with time-dependent density-functional theory (TDDFT) or other ab initio methods.
The purpose of the PHM is to give detailed insight into electronic excitation processes which is not obtainable
from local visualization methods such as transition densities, density differences, or natural transition orbitals.
The PHM is defined as a nonlocal function of two spatial variables and provides information about the origins,
destinations, and connections of charge fluctuations during an excitation process; it is particularly valuable
to analyze charge-transfer excitonic processes. In contrast with the transition density matrix,
the PHM has a statistical interpretation involving joint probabilities of individual states and their transitions,
it satisfies several sum rules and exact conditions, and it is easier to read and interpret.
We discuss and illustrate the properties of the PHM and give several examples and applications to excitations in
one-dimensional model systems, in a hydrogen chain, and in a benzothiadiazole based molecule.
\end{abstract}

\section{Introduction}\label{sec1}

The first-principles calculation of electronic excitation energies and excited-state properties is among the most
important tasks in quantum chemistry and many other areas of science. Time-dependent density-functional theory (TDDFT)
\cite{Runge1984,Ullrich2012,Ullrich2014} has achieved much popularity for calculating excitation energies and for
simulating electron dynamics in real time, thanks to its favorable balance of accuracy and computational efficiency.
TDDFT yields excited-state properties such as transition frequencies, oscillator strengths, or excited-state
forces and geometries that are of similar quality as results obtained in ground-state DFT.\cite{Casida2012,Adamo2013}

Within a single-particle picture such as in TDDFT, molecular electronic excitations can be analyzed to determine which
single-particle transition is dominant; at times, several single-particle transitions contribute significantly
to the excitation of the many-body system. However, it is often desirable to gain additional insight into the nature of an
electronic transition, and to analyze the
change in the many-body wave function for specific patterns in the rearrangement of the electronic configuration. This is very useful, for instance,
in the treatment of charge-transfer excitations, or for conjugated donor-acceptor molecules. There exist various tools for the analysis
of electronic transitions, as reviewed by Dreuw and Head-Gordon.\cite{Dreuw2005}
First of all, one can simply inspect the
relevant molecular orbitals close to the highest occupied and lowest unoccupied molecular orbital
(HOMO-1, HOMO, LUMO, LUMO+1, etc.).\cite{Dreuw2004,Rappoport2004} If one transition, say HOMO-LUMO, is dominant,
such analysis will be straightforward.  Likewise, the density difference between
the excited state and the ground state often provides useful insights.\cite{Sun2005,Stein2009}
Other widely used tools to analyze electronic excitations in molecules are transition densities\cite{Beenken2004,Jespersen2004}
and natural transition orbitals.\cite{Martin2003,Korzdorfer2009,Richard2011,Nitta2012}

The analysis tools mentioned so far all have in common that they are local (i.e., they depend on a single position vector $\bfr$).
There are, however, important aspects of electronic excitations that cannot be described using a local quantity,
but are intrinsically nonlocal (i.e. they depend on two position vectors, $\bfr$ and $\bfr'$). For instance, an exciton is
a bound electron-hole pair, and one should ask about the position of both the electron and the hole. Or, for a charge-transfer
excitation process, one might want to know which parts of the system act as donors of charge, and where that charge is transferred to;
in other words, one would like a map that indicates origins and destinations of charge transfer in a molecule or molecular complex,
and which donor and acceptor regions are connected. The
density difference or the transition density do not contain enough information to answer these questions.

The transition density matrix (TDM) \cite{McWeeny1960} is a well-known concept in quantum chemistry,
which has been widely used as a nonlocal visualization tool to map excitonic effects
and electron-hole pair coherences for molecular excitations.
\cite{Mukamel1997,Tretiak1998,Tretiak2002,Tretiak2003,Tretiak2005,Wu2006,Igumenshchev2007,Li2007,Kilina2008,Wong2009,Xia2012,Bappler2014,Plasser2014,Etienne2015}
The TDM is defined as the density matrix connecting the molecular ground state and a particular excited state
(the precise definition will be given below in Section \ref{sec3.3}).
However, in spite of its popularity, the TDM suffers from various shortcomings as we will make clear
in the following. The purpose of this paper is to introduce an alternative to it, the particle-hole map (PHM).

Before giving precise definitions of the PHM and the TDM (see Section 3), we will provide some motivation via a simple
illustration. We consider a one-dimensional (1D) model, described by the Kohn-Sham equation
\begin{equation}\label{eq1.1}
\left[-\frac{\hbar^{2}}{2m}\frac{d^{2}}{dx^{2}}+\frac{e^2}{4\pi\epsilon_0}\int\frac{n(x')dx'}{\sqrt{(x-x')^2 + \alpha^2}}+v_{\rm ext}(x)\right]\varphi_{i}(x)=\varepsilon_{i}\varphi_{i}(x).
\end{equation}
Here, $\alpha$ is a parameter to regularize the 1D Coulomb interaction, and
$v_{\rm ext}(x)$ is the external confining potential; for simplicity, we only include the Hartree potential and set the exchange-correlation (xc)
potential to zero. The occupied and unoccupied Kohn-Sham eigenfunctions $\varphi_i(x)$ serve as input to a TDDFT calculation to obtain the excitation spectrum and the PHM or TDM corresponding to the $n$th excitation.

\begin{figure}[t]
\centering
\includegraphics[width=0.95\linewidth]{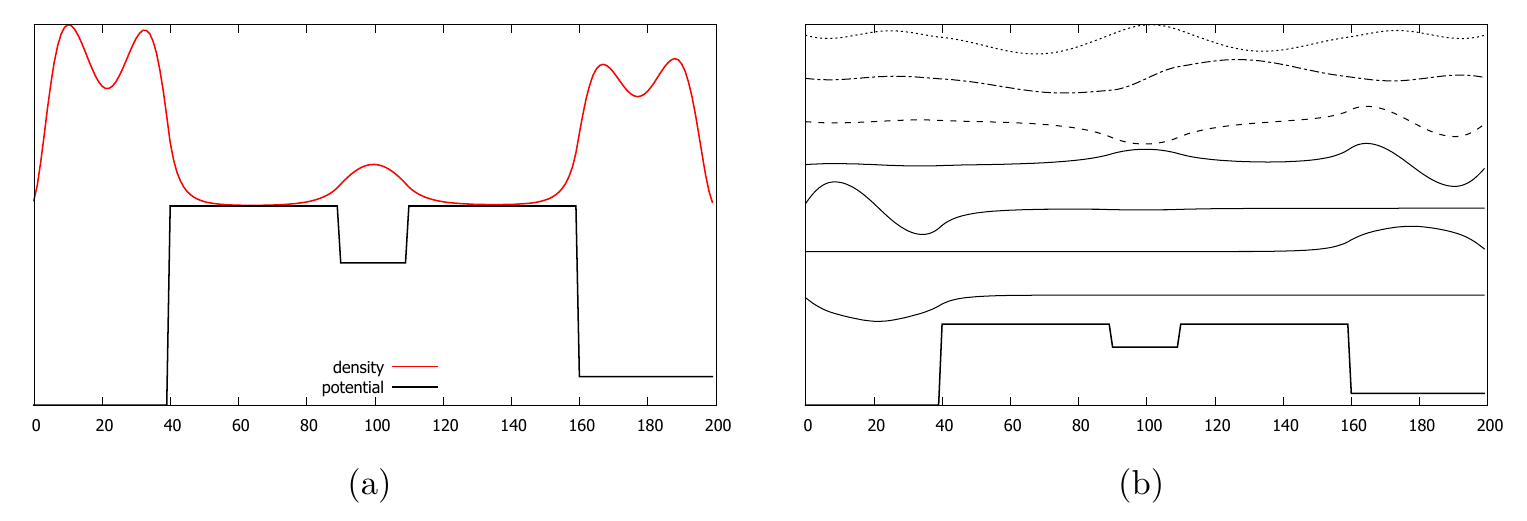}
\caption{1D model with 8 electrons confined in a triple potential well. (a) ground-state density and potential wells, (b) lowest 7 single-particle wave functions.} \label{fig:1DQWs}
\end{figure}

We discretize Eq. (\ref{eq1.1}) on an equidistant lattice with 200 points in a box with length 10 a.u. and infinite walls. As the left panel of
Fig. \ref{fig:1DQWs} shows, $v_{\rm ext}(x)$ is slightly asymmetric: the left well is a bit deeper than the right one.
The system has 8 electrons and hence 4 doubly occupied orbitals. The ground-state density indicates that the
electrons mainly stay in the left and right wells. Figure \ref{fig:1DQWs} also plots the first 7 single-particle eigenstates; clearly,
the confinement from the wells becomes less important for higher levels.

\begin{figure}[t]
\centering
\includegraphics[width=0.95\linewidth]{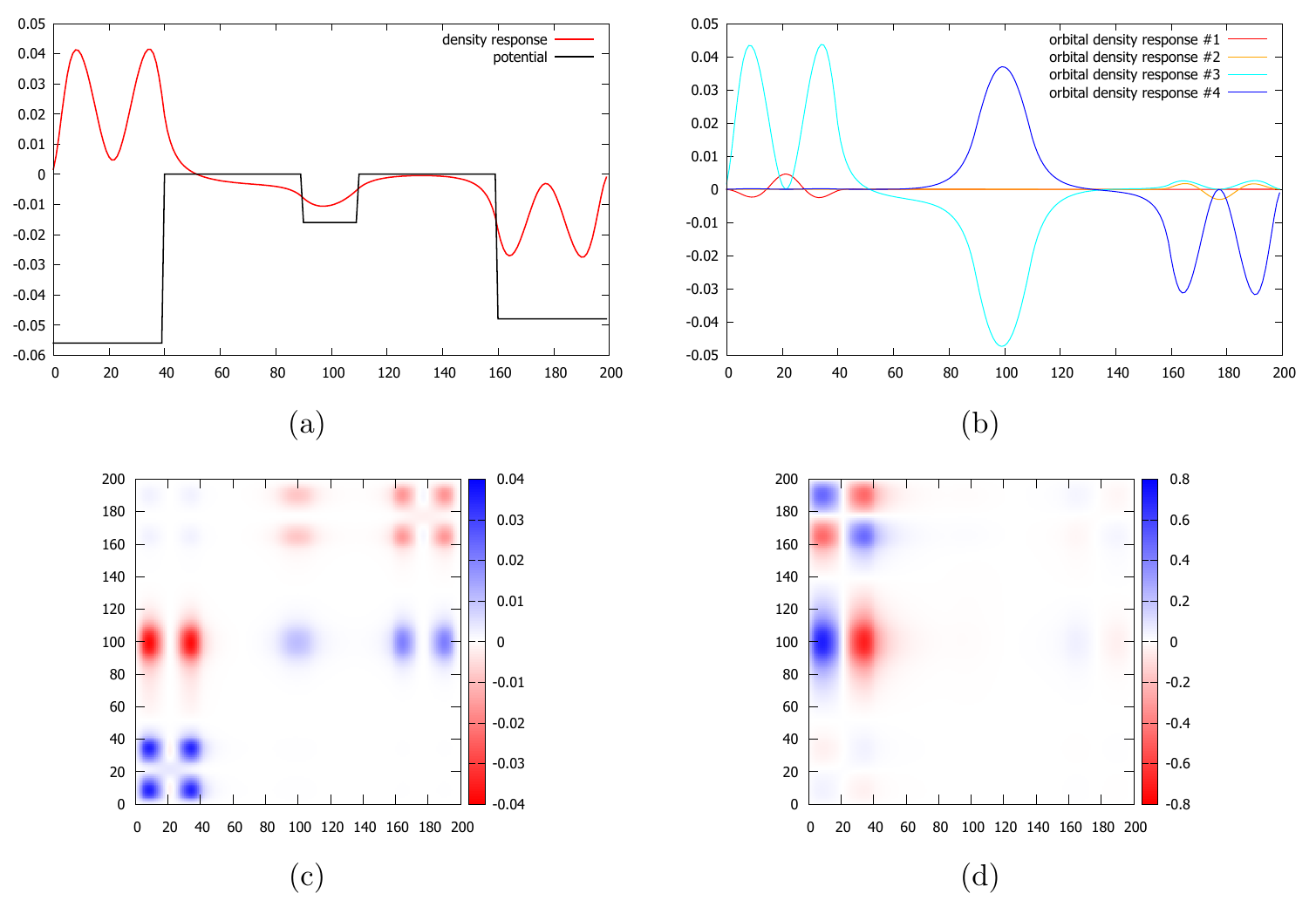}
\caption{(a) density response associated with the first excitation;
(b) individual density responses of the four occupied orbitals;
(c) associated Particle-Hole Map;
(d) associated Transition Density Matrix.} \label{fig:1DQWs_ex1}
\end{figure}

Figure \ref{fig:1DQWs_ex1} shows the total density response, the orbital density responses, the PHM and the TDM for the lowest excitation. The density response indicates an electron transfer between the left and the right wells. However, the individual orbital density responses reveal that the smaller well in the
middle of the system plays a very important role: the responses of the third and fourth orbital have a very large peak in the middle, which cancels
out in the total response. Thus, even in a simple 1D system like this one, the problem of identifying the internal motion of the individual charges during an
excitation process is not straightforward.

Unfortunately, the TDM does not provide much help in the analysis of the excitation process, but instead leads to a rather distorted picture.
As the lower right panel of Fig. \ref{fig:1DQWs_ex1} shows, the TDM does give a hint of the importance of the small well in the middle,
but it produces an asymmetric picture which is rather misleading.

Instead, the PHM (lower left panel of Fig. \ref{fig:1DQWs_ex1}) provides a very clear picture of the charge redistribution during the excitation process:
its horizontal axis indicates where charges are coming from, and the vertical axis tells us where they are going to; red and blue indicate electrons
and holes, respectively. One can see very nicely that the charge transfer from left to right proceeds via the middle well.
We will justify this interpretation of the PHM below, and will outline a set of simple rules how to construct and how to read it.

The remainder of this paper is organized as follows. In Section \ref{sec2} we give a brief summary of the
basic elements of TDDFT which will be needed further on. Section \ref{sec:PHM} contains the definition of the PHM
in the time domain and in the linear-response frequency domain, as well as a comparison with the TDM.
Section \ref{sec4} contains examples and discussions: in Section \ref{sec4.1} we return to our 1D model system,
and in Section \ref{subsec:hydrogen_chain} and \ref{subsec:bt} we discuss application to a Hydrogen chain and a Benzothiadiazole based molecule. Section \ref{sec5} gives our conclusions.

\section{A brief summary of TDDFT}\label{sec2}

In this Section, we briefly summarize the essentials of TDDFT which will be needed for the discussion later on.
More details on the formal framework of TDDFT can be found elsewhere.\cite{Ullrich2012,Ullrich2014}

In the following, we will consider $N$-electron systems which are
not magnetic or spin polarized so that we can ignore the spin index. The systems are initially in
the ground state, described by the static Kohn-Sham equation
\begin{equation}\label{eq2.1}
\left[ -\frac{\hbar^2 \nabla^2}{2m} + v_{\rm nuc}(\bfr) + \frac{e^2}{4\pi\epsilon_0}\int d^3r' \: \frac{n_0(\bfr')}{|\bfr - \bfr'|}
+ v_{\rm xc}^{(0)}[n_0](\bfr)\right]\varphi^{(0)}_j(\bfr) = \varepsilon_j \varphi^{(0)}_j(\bfr) \:.
\end{equation}
Here, $v_{\rm nuc}(\bfr)$ is the electrostatic potential caused by fixed classical nuclei (we make the Born-Oppenheimer approximation), and
$v_{\rm xc}^{(0)}[n_0](\bfr)$ is the xc potential of ground-state DFT, which is a functional
of the ground-state density
\begin{equation}\label{eq2.2}
n_0(\bfr) = \sum_{j=1}^N |\varphi_j^{(0)}(\bfr)|^2 \:.
\end{equation}
Equation (\ref{eq2.1}) is solved self-consistently, using a suitable approximation for $v_{\rm xc}^{(0)}[n_0](\bfr)$.

The majority of applications of TDDFT fall into two categories: propagation in real time under arbitrary external perturbations,
or calculation of the frequency-dependent linear-response, which allows one to extract, among other things, the excitation spectrum of the system.
In this paper we will consider both situations.

The time-dependent Kohn-Sham equation,
\begin{equation}\label{eq2.3}
\left[ -\frac{\hbar^2 \nabla^2}{2m} + v_{\rm nuc}(\bfr) + v_{\rm ext}(\bfr,t) + \frac{e^2}{4\pi\epsilon_0}\int d^3r' \: \frac{n(\bfr',t)}{|\bfr - \bfr'|}
+ v_{\rm xc}[n](\bfr,t)
\right]\varphi_j(\bfr,t) = i \hbar \frac{\partial}{\partial t} \,  \varphi_j(\bfr,t) \:,
\end{equation}
propagates a set of $N$ Kohn-Sham orbitals forward in time, under the influence of an arbitrary time-dependent
external potential $v_{\rm ext}(\bfr,t)$ that is switched on at the initial time $t_0$. Similar to ground-state DFT,
the time-dependent xc potential $v_{\rm xc}[n](\bfr,t)$ is a functional of the time-dependent density
\begin{equation}\label{eq2.4}
n(\bfr,t) = \sum_{j=1}^N |\varphi_j(\bfr,t)|^2
\end{equation}
and needs to be approximated in practice. Most commonly used is the adiabatic approximation $v_{\rm xc}^{\rm adia}[n](\bfr,t)
= \left. v_{\rm xc}^{(0)}[\tilde n](\bfr)\right|_{\tilde n = n(\bfr,t)}$. The time-dependent Kohn-Sham equation (\ref{eq2.3})
is an initial value problem, where $\varphi_j(\bfr,t_0) = \varphi^{(0)}_j(\bfr)$. From the time-dependent density $n(\bfr,t)$
one then calculates the physical observables of interest such as, for instance, the dipole moment ${\bf d}(t) = \int d^3r \: \bfr n(\bfr,t)$
and the spectral properties following from it. In this way one can calculate the optical spectra of molecules, following
a pulsed excitation and subsequent free time propagation.\cite{Yabana2006}

A more widely used approach for calculating excitation energies and excited-state properties of molecules is
via the frequency-dependent linear response. The idea is that an electronic excitation corresponds to an eigenmode
of the electronic density, which can oscillate without any driving force, similar to the normal modes of a classical
system of coupled oscillators. These electronic eigenmodes are found via solution of the so-called Casida equation:\cite{Casida1995}
\begin{equation} \label{eq2.5}
\left( \begin{array}{cc} \underline{\bf A} & \underline{\bf K} \\ \underline{\bf K} & \underline{\bf A} \end{array} \right)
\left( \begin{array}{c} {\bf X} \\ {\bf Y} \end{array} \right)
= \Omega
\left( \begin{array}{cc} -\underline{\bf 1} & \underline{\bf 0} \\ \underline{\bf 0} & \underline{\bf 1} \end{array} \right)
\left( \begin{array}{c} {\bf X} \\ {\bf Y} \end{array} \right),
\end{equation}
where the elements of the matrices $\underline{\bf A}$ and $\underline{\bf K}$ are given by
\begin{eqnarray} \label{eq2.6}
A_{ia,i'a'}(\omega) &=& \delta _{ii'} \delta_{aa'} \omega_{ai} +
K_{ia,i'a'}(\omega)
\\
K_{ia,i'a'}(\omega) &=&\int d^3r \int d^3r' \varphi_i^{(0)*}(\bfr) \varphi_a^{(0)}(\bfr)\left[ \frac{1}{|\bfr-\bfr'|} +
f_{\rm xc}(\bfr,\bfr',\omega)\right] \varphi_{i'}^{(0)}(\bfr')\varphi_{a'}^{(0)*}(\bfr') \label{eq2.7}
\end{eqnarray}
and $i,i'$ and $a,a'$ run over occupied and unoccupied Kohn-Sham orbitals, respectively. The xc kernel
$f_{\rm xc}(\bfr,\bfr',\omega)$ is defined via the Fourier transform of
\begin{equation}\label{eq2.8}
f_{\rm xc}(\bfr,t,\bfr',t') = \left. \frac{\delta v_{\rm xc}[n](\bfr,t)}{\delta n(\bfr',t')}\right|_{n_0(\bfr)} \:.
\end{equation}
Almost all commonly used expressions for $f_{\rm xc}$ are constructed within the adiabatic approximation, wherein the frequency dependence
is ignored.

Solution of Eq. (\ref{eq2.5}) gives, in principle, the exact excitation energies of the $n$th excited state $\Omega_n$.
The eigenvectors $\bf X$ and $\bf Y$ can be used to calculate the related oscillator strengths as well as the
transition densities
\begin{equation}\label{eq2.9}
n_1(\bfr,\Omega_n) = \sum_{ia}\left[ \varphi_i^{(0)}(\bfr) \varphi_a^{(0)*}(\bfr) X_{ia}(\Omega_n)
+ \varphi_i^{(0)*}(\bfr) \varphi_a^{(0)}(\bfr) Y_{ia}(\Omega_n)\right],
\end{equation}
i.e., the density fluctuations associated with the $n$th electronic eigenmode.

For real orbitals, the Casida equation (\ref{eq2.5}) can be recast into the simpler form
\begin{equation}\label{eq2.10}
\underline{\bf C}{\bf Z} = \Omega^2 {\bf Z} \:,
\end{equation}
where $\underline{\bf C} = (\underline{\bf A} - \underline{\bf B})^{1/2}(\underline{\bf A}+\underline{\bf B})
(\underline{\bf A} - \underline{\bf B})^{1/2}$ and ${\bf Z}=(\underline{\bf A} - \underline{\bf B})^{1/2}
({\bf X}-{\bf Y})$. The transition densities then become
\begin{equation}\label{eq2.11}
n_1(\bfr,\Omega_n) = \sum_{ia}\varphi_i^{(0)}(\bfr) \varphi_a^{(0)}(\bfr)
\omega_{ia}^{1/2} Z_{ia}(\Omega_n) \:.
\end{equation}

\section{Definition and properties of the Particle-Hole Map}\label{sec:PHM}

\subsection{Densities and orbital densities and their first-order responses}\label{subsec:GeneralPHM}

Let us consider an $N$-electron system, starting at time $t_0$ from the Kohn-Sham ground state, and evolving in time
under the influence of an external time-dependent potential.
The time-dependent Kohn-Sham orbitals can be written as
\begin{equation}
\varphi_i(\bfr,t) = \varphi_i^{(0)}(\bfr)e^{-i\varepsilon_i t/\hbar} + \delta \varphi_i(\bfr,t)e^{-i\varepsilon_i t/\hbar} \:,
\qquad i=1,\ldots, N\:,
\end{equation}
where $t\ge t_0$ and  $\delta\varphi_i(\bfr,t)$ denotes the deviation of the $i$th Kohn-Sham orbital from its initial form,
which can be expanded as
\begin{equation}\label{expansion}
\delta \varphi_i(\bfr,t) = \sum_{k=1}^\infty C_{ik}(t) \varphi_k^{(0)}(\bfr) e^{i\omega_{ik}t} \:,
\end{equation}
where $\omega_{ik} = (\epsilon_i - \epsilon_k)/\hbar$.
The expansion coefficients can be calculated using standard time-dependent perturbation theory: to first order  one finds
\begin{equation}
 C_{ik}^{(1)}(t) = \frac{1}{i\hbar}\int_{t_0}^t H'_{ki}(t')e^{i\omega_{ki} t'} dt' \:,
\end{equation}
where $H'_{ki}(t)$ is the matrix element of the time-dependent Kohn-Sham potential
between the states $\varphi_k^{(0)}(\bfr)$ and $\varphi_i^{(0)}(\bfr)$. In the following, we assume that
all Kohn-Sham ground-state orbitals are real.

The expansion (\ref{expansion}) involves a sum over occupied and unoccupied Kohn-Sham orbitals, and we can
split it up as $\delta \varphi_i(\bfr,t) = \delta \varphi_{i,o}(\bfr,t) + \delta \varphi_{i,u}(\bfr,t)$, where
\begin{equation}
\delta \varphi_{i,o}(\bfr,t) = \sum_{j=1}^N C_{ij}(t) \varphi_j^{(0)}(\bfr) e^{i\omega_{ij} t} \:, \qquad
\delta \varphi_{i,u}(\bfr,t) = \sum_{a=N+1}^\infty C_{ia}(t) \varphi_a^{(0)}(\bfr) e^{i\omega_{ia} t}  \:.
\end{equation}
We can also obtain $\delta \varphi_{i,u}(\bfr,t)$ by projection as
\begin{equation}
\delta \varphi_{i,u}(\bfr,t) = \varphi_i(\bfr,t) - \sum_{j=1}^N \varphi_j^{(0)}(\bfr)\int d^3r' \varphi_i(\bfr',t)\varphi_j^{(0)}(\bfr') \:.
\end{equation}
The time-dependent density is given by
\begin{equation}
n(\bfr,t) = \sum_{i=1}^N |\varphi_i(\bfr,t)|^2  = \sum_{i=1}^N n_i(\bfr,t) \:,
\end{equation}
which defines the orbital densities $n_i(\bfr,t)$. From the perturbation expansion of the orbitals we obtain the first-order
orbital densities as
\begin{equation}
n_i^{(1)}(\bfr,t)  = n_{i,o}^{(1)}(\bfr,t) + n_{i,u}^{(1)}(\bfr,t) \:,
\end{equation}
where
\begin{equation}
n_{i,o}^{(1)}(\bfr,t) =
\sum_{j=1}^N \varphi_i^{(0)}(\bfr)\varphi_j^{(0)}(\bfr)
\left[e^{i\omega_{ij} t}C_{ij}^{(1)}(t) + e^{i\omega_{ji}t}C_{ij}^{(1)*}(t) \right]
\end{equation}
and
\begin{equation}
n_{i,u}^{(1)}(\bfr,t) =
\sum_{a=N+1}^\infty \varphi_i^{(0)}(\bfr)\varphi_a^{(0)}(\bfr)
\left[e^{i\omega_{ia} t} C_{ia}^{(1)}(t) + e^{i\omega_{ai}t} C_{ia}^{(1)*}(t) \right].
\end{equation}
It is straightforward to see that the total first-order density response is given by
\begin{equation}
n_1(\bfr,t) = \sum_{i=1}^N n_{i,u}^{(1)}(\bfr,t) \:,
\end{equation}
and all contributions from $n_{i,o}^{(1)}(\bfr,t)$ sum up to zero. In other words, the total density response
only comes from transitions to unoccupied orbitals, as expected. Rotations within the space of occupied
Kohn-Sham orbitals are not reflected in the total density response, but they do contribute to the orbital density responses
via $n_{i,o}^{(1)}(\bfr,t)$. These contributions can be significant, and they will be discarded in constructing the PHM.

\subsection{The time-dependent particle-hole map}\label{sec3.1}
Let us consider the following object:
\begin{equation}\label{eq3.0}
P(\bfr,\bfr',t) =\sum_{i=1}^N \left\{ |\varphi_i(\bfr',t)|^2 - |\varphi_i^{(0)}(\bfr')|^2\right\}|\varphi_i^{(0)}(\bfr)|^2 \:.
\end{equation}
This expression has a straightforward interpretation: it is the sum of the density fluctuations of each time-evolving Kohn-Sham
orbital, $|\varphi_i(\bfr',t)|^2 - |\varphi_i^{(0)}(\bfr')|^2$, weighted by the density of the $i$th ground-state orbital
$|\varphi_i^{(0)}(\bfr)|^2$. Each term in the sum has the form of a difference of joint probabilities: it is the product of the probability
density of the $i$th ground-state Kohn-Sham orbital at position $\bfr$, with the probability density fluctuation, relative to the
ground state, of the $i$th orbital at time $t$ and position $\bfr'$.

Let us now assume that the time-dependent perturbation is sufficiently small so that we can neglect terms  of order $\delta^2$.
We then obtain
\begin{eqnarray}
|\varphi_i(\bfr',t)|^2 - |\varphi_i^{(0)}(\bfr')|^2
&=& \varphi_i^{(0)}(\bfr')  \delta \varphi_i(\bfr',t) + c.c.
\nonumber\\
&=& \varphi_i^{(0)}(\bfr')  \delta \varphi_{i,o}(\bfr',t)
+ \varphi_i^{(0)}(\bfr')  \delta \varphi_{i,u}(\bfr',t) + c.c.
\end{eqnarray}
We now define the PHM as follows:
\begin{equation}\label{eq3.1}
\Xi(\bfr,\bfr',t) = \sum_{i=1}^N \left\{\varphi_i^{(0)}(\bfr')  \delta \varphi_{i,u}(\bfr',t) + c.c. \right\}|\varphi_i^{(0)}(\bfr)|^2 \:.
\end{equation}
The PHM thus involves only those orbital density fluctuations that result from transitions into initially unoccupied Kohn-Sham states.
In other words, $\Xi(\bfr,\bfr',t)$ is the sum of joint probabilities
that a particle originates at position $\bfr$ and moves, during the excitation process, to position $\bfr'$.
We will illustrate this later on with several examples.

It must be pointed out that $\Xi(\bfr,\bfr',t)$ is expressed in terms of the Kohn-Sham orbitals, which in and by themselves do not
have a rigorous physical meaning (it is in principle possible to construct a PHM using time-dependent natural
orbitals,\cite{Pernal2007,Appel2010,Dutoi2014} which may have some formal advantages). Nevertheless,
the time-dependent PHM has several exact properties. First of all, it vanishes at the initial time, $\Xi(\bfr,\bfr',t_0)=0$,
and it also vanishes for $t>t_0$ if $v_{\rm ext}(\bfr,t)$ is constant (of course provided the system starts from the ground state).
In addition to these rather obvious requirements, the PHM satisfies two important sum rules:
\begin{equation}\label{eq3.2}
\int \Xi(\bfr,\bfr',t) d^3r'= 0
\end{equation}
and
\begin{equation}\label{eq3.3}
\int \Xi(\bfr,\bfr',t) d^3r= n_1(\bfr',t) \:.
\end{equation}
Equation (\ref{eq3.2}) is due to norm conservation; in the present context it expresses the fact that
if during the excitation process an electron is emitted from position $\bfr$, a hole of equal and opposite charge remains behind,
so that the overall charge of the system does not change.
Equation (\ref{eq3.3}), on the other hand, shows that if we integrate over all the ``origins'', the PHM simply delivers
the linear density response $n_1(\bfr',t)$, which is obtained in principle exactly from TDDFT.

\subsection{The frequency-dependent particle-hole map}\label{sec3.2}

The time-dependent PHM, Eq. (\ref{eq3.1}), provides a real-time visualization of the particle-hole dynamics of electronic excitations,
following the solution of the time-dependent Kohn-Sham equation (\ref{eq2.3}). Most applications of TDDFT, however, are
in the frequency-dependent linear-response regime. Let us derive the associated frequency-dependent PHM, which targets
a specific excitation with energy $\Omega_n$.

Inserting the expansion (16) into Eq. (\ref{eq3.1}) we obtain
\begin{equation} \label{eq3.7}
\Xi(\bfr,\bfr',t) =
\sum_{i=1}^N\sum_{a=N+1}^\infty\left\{ C_{ia}(t)e^{i\omega_{ia}t}\varphi_i^{(0)*}(\bfr') \varphi_a^{(0)}(\bfr') + c.c. \right\}|\varphi^{(0)}_i(\bfr)|^2 \:.
\end{equation}
To identify the coefficients $C_{ia}(t)$, we consider the linear-response case and make use of the sum rule (\ref{eq3.3}).
If the system is in an electronic eigenmode corresponding to the $n$th excited state, then the linear density response is given by
\begin{equation}\label{eq3.8}
n_1(\bfr,t) = e^{-i\Omega_n t}n_1(\bfr,\Omega_n) \:.
\end{equation}
But the frequency-dependent density response follows from the Casida equation, see Eq. (\ref{eq2.9}), and comparison suggests that
the frequency-dependent PHM is given by
\begin{equation}\label{eq3.9}
\Xi(\bfr,\bfr',\Omega_n) = \sum_{ia} \left\{ \varphi_i^{(0)}(\bfr')\varphi_a^{(0)*}(\bfr')X_{ia}(\Omega_n)
+ \varphi_i^{(0)*}(\bfr')\varphi_a^{(0)}(\bfr')Y_{ia}(\Omega_n)\right\} |\varphi^{(0)}_i(\bfr)|^2 \:.
\end{equation}
For the case of real orbitals, see Eqs. (\ref{eq2.10}) and (\ref{eq2.11}), this expression simplifies to
\begin{equation}\label{eq3.10}
\Xi(\bfr,\bfr',\Omega_n) = \sum_{ia}|\varphi^{(0)}_i(\bfr)|^2 \varphi_i^{(0)}(\bfr') \varphi_a^{(0)}(\bfr')
\omega_{ia}^{1/2} Z_{ia}(\Omega_n) \:.
\end{equation}
Thus, the frequency-dependent PHM can be obtained in a straightforward manner from the solutions of the Casida equation.
It is easy to see that it satisfies the sum rules
\begin{equation}\label{eq3.11}
\int \Xi(\bfr,\bfr',\Omega_n) d^3r'= 0\:, \qquad \qquad
\int \Xi(\bfr,\bfr',\Omega_n) d^3r= n_1(\bfr',\Omega_n) \:.
\end{equation}
in analogy with the sum rules (\ref{eq3.2}) and (\ref{eq3.3}) for the time-dependent PHM.

In the following, we use the abbreviations PHM$_\omega$ and PHM$_t$ for the frequency-dependent and time-dependent particle-hole maps, respectively.

\subsection{Comparison with the transition density matrix}\label{sec3.3}

The TDM associated with an electronic transition between the many-body ground state $\Psi_0$ and the $n$th excited state $\Psi_n$
is defined as\cite{McWeeny1960,Tretiak2002,Furche2001}
\begin{equation}\label{eq3.12}
\Gamma_n(\bfr,\bfr') = \langle \Psi_n | \hat \rho(\bfr,\bfr') | \Psi_0\rangle \:,
\end{equation}
where $\hat \rho(\bfr,\bfr')$ is the reduced one-particle density matrix operator. The TDM can be obtained
from the solutions of Eq. (\ref{eq2.5}) as follows:\cite{Furche2001}
\begin{equation}\label{eq3.13}
\Gamma_n(\bfr,\bfr') = \sum_{ia} \left\{ \varphi_i^{(0)}(\bfr')\varphi_a^{(0)*}(\bfr)X_{ia}(\Omega_n)
+ \varphi_i^{(0)*}(\bfr)\varphi_a^{(0)}(\bfr')Y_{ia}(\Omega_n)\right\}.
\end{equation}
For real orbitals, see Eqs. (\ref{eq2.10}) and (\ref{eq2.11}), this becomes
\begin{equation}\label{eq3.14}
\Gamma_n(\bfr,\bfr') = \frac{1}{2}\sum_{ia} \left\{ \varphi_i^{(0)}(\bfr')\varphi_a^{(0)}(\bfr)(\omega^{1/2}_{ai} + \omega^{-1/2}_{ai})
+ \varphi_i^{(0)}(\bfr)\varphi_a^{(0)}(\bfr')(\omega^{1/2}_{ai} - \omega^{-1/2}_{ai})\right\}Z_{ia}(\Omega_n) \:.
\end{equation}
The diagonal of the TDM is equal to the transition density, i.e., $\Gamma_n(\bfr,\bfr)=n_1(\bfr,\Omega_n)$.

In an earlier paper, we defined the time-dependent TDM as\cite{Li2011}
\begin{equation}\label{eq3.15}
\Gamma(\bfr,\bfr',t) = \langle \Psi(t) | \hat \rho(\bfr,\bfr') | \Psi_0 e^{-iE_0t}\rangle \:,
\end{equation}
where $\Psi(t)$ is the time-dependent many-body wave function that evolves from the ground state $\Psi_0$.
In TDDFT, the exact wave function is not available, but it can be approximated by the time-dependent Kohn-Sham
Slater determinant. The time-dependent TDM can then be expressed in terms of the time-dependent Kohn-Sham orbitals.
One can then plot the difference of the absolute squares of the time-dependent TDM and the ground-state Kohn-Sham density matrix,
$|\Gamma^{\rm KS}(\bfr,\bfr',t)|^2 - |\Gamma_0^{\rm KS}(\bfr,\bfr')|^2$.

\section{Implementation and examples}\label{sec4}

\subsection{1D model system revisited: how to read the PHM}\label{sec4.1}

The PHM produces a map which illustrates the charge motion and redistribution during an excitation process.
There are various ways in which such a map can be realized in practice.
For example, in our 1D triple well model introduced in Section \ref{sec1}, the full PHM$_\omega$, $\Xi(x,x',\Omega_n)$, has $200 \times 200$ pixels
(corresponding to the 200 grid points). The horizontal label (the $x$ axis) indicates where an electron or hole is originating from,
and the vertical label (the $x'$ axis) indicates where it is going to. Each pixel contains the associated joint probability fluctuation,
according to definition (\ref{eq3.1}). This can be a positive or a negative number: we associate a positive number with the probability of an electron
(here, color coded as blue) and a negative number with the probability of a hole (represented as red).
Any signature in the PHM which is located near the diagonal indicates that origin and
destination coincide, so there is no transfer.

Let us now give a more detailed explanation of the PHM$_\omega$ shown in Fig. \ref{fig:1DQWs_ex1}.
From the orbital density response we can see that there are two major contributions in the first excitation, the
left-middle and the right-middle charge transfer (CT). Column 1-40 contain the information of left-middle CT since they
describe the excitations {\em from} the left well. The four blue patches indicate an electron staying in the left quantum well
while a hole goes to the middle well. The detailed structure is related to the double peaks of the ground state density on the left and on the right.

From the vertical band around column 100, we see that an electron is left in the middle well and the hole goes to the right well. Similarly, we can read the vertical band between columns 160 to 200 as representing an electron transfer to the middle well and a corresponding hole left behind.
The horizontal rows represent how the final density response is a superposition of different CT processes. For example, the horizontal band in the middle of the map shows how the hole from the right well is neutralized by the electron from center and right wells. For each point in the system, summing over all the electron and hole probabilities coming from the whole system leads to the total density response.

By contrast, the TDM is difficult to interpret. The diagonal, which is the density response, is relatively small compared to the left top block. Mathematically, it is due to the fact that orbital 3 and orbital 5 contain nearly no common non-zero part. On this level, one does not see a clear explanation of the observed
pattern, and one cannot derive any useful information from it. It is also difficult to apply the delocalization length and the coherence length concept to the plot since this is not a large molecular chain system.\cite{Tretiak2002}

\begin{figure}[t]
\centering
\includegraphics[width=0.95\linewidth]{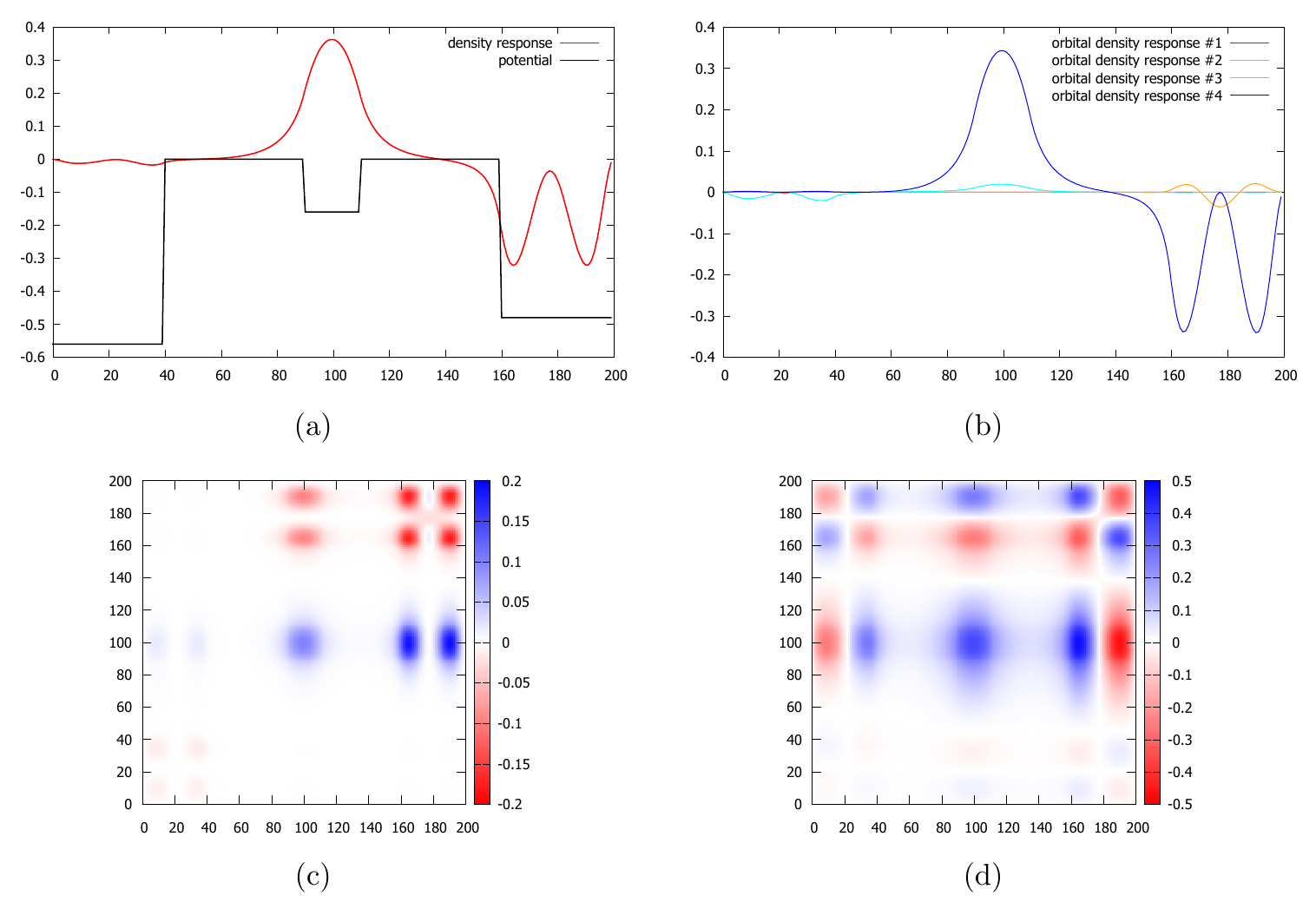}
\caption{(a) density response associated with the second excitation;
(b) individual density responses of the four occupied orbitals;
(c) associated PHM$_\omega$;
(d) associated TDM.} \label{fig:1DQWs_ex2}
\end{figure}

\begin{figure}[t]
\centering
\includegraphics[width=0.95\linewidth]{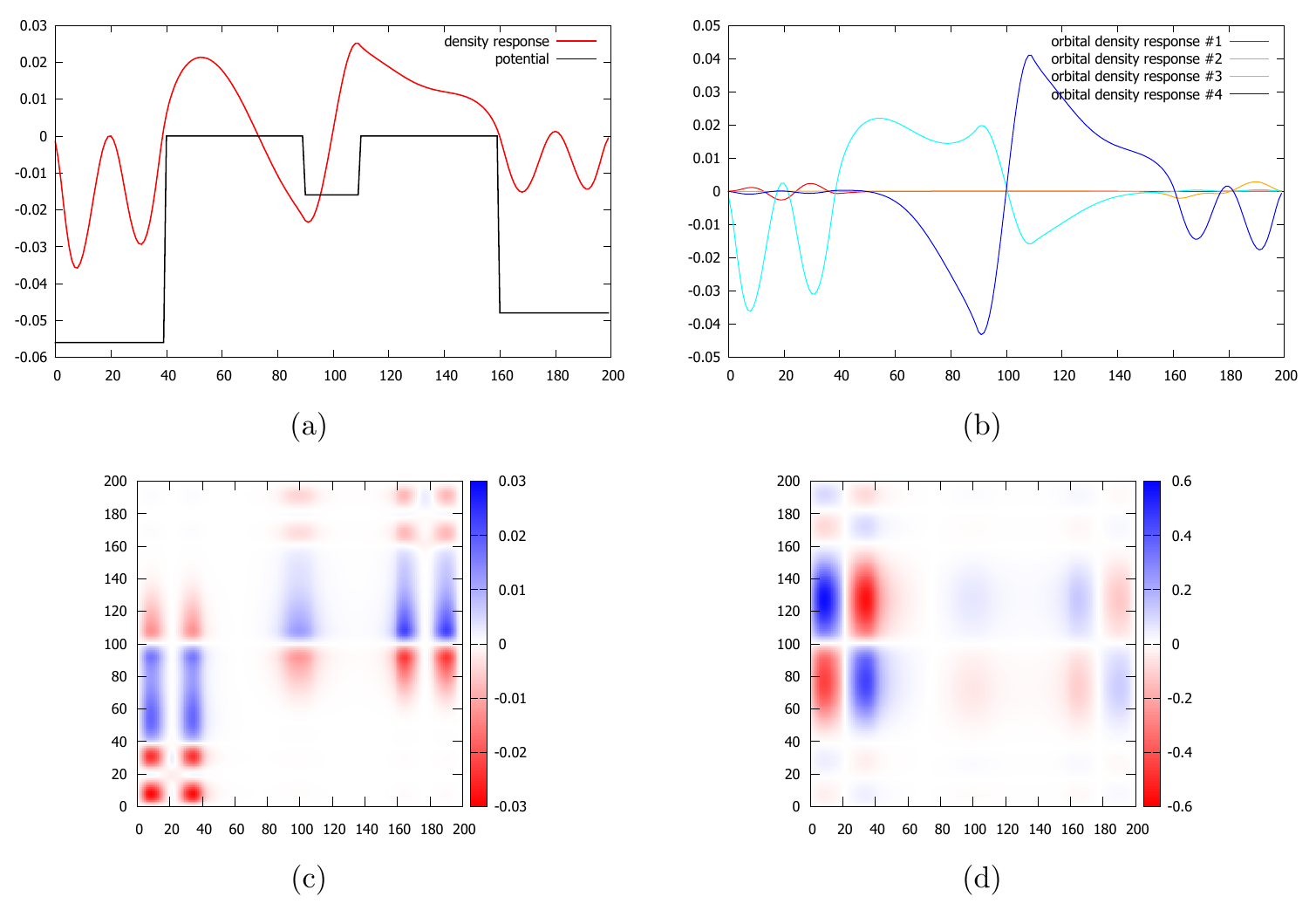}
\caption{As in Fig. \ref{fig:1DQWs_ex2}, but for the third excitation.} \label{fig:1DQWs_ex3}
\end{figure}

In Fig. \ref{fig:1DQWs_ex2}, which illustrates the second excitation, the PHM$_\omega$ and the TDM are not so drastically different. From the PHM$_\omega$, we see that the upper right block (80-200) is similar to what was seen for the first excitation. It shows that this excitation only involves charge fluctuations within the right part of the system. The TDM gives a similar message, but one needs to reckon with respect to the diagonal, which is not so intuitive.

For the third excitation, see Fig. \ref{fig:1DQWs_ex3}, the charge redistributions are no longer confined to the wells but also appear on top of the barriers. The PHM$_\omega$ shows that the charge fluctuation on the left barrier comes from the left well only, while the charge on the right barrier comes from the middle and right wells. The signatures on the barriers are rather diffuse. As before in Fig. 2, the TDM suffers from a lack of clarity along the diagonal.

\subsection{1D model system revisited: PHM$_\omega$ versus PHM$_t$}

\begin{figure}[t]
\centering
\includegraphics[width=0.95\linewidth]{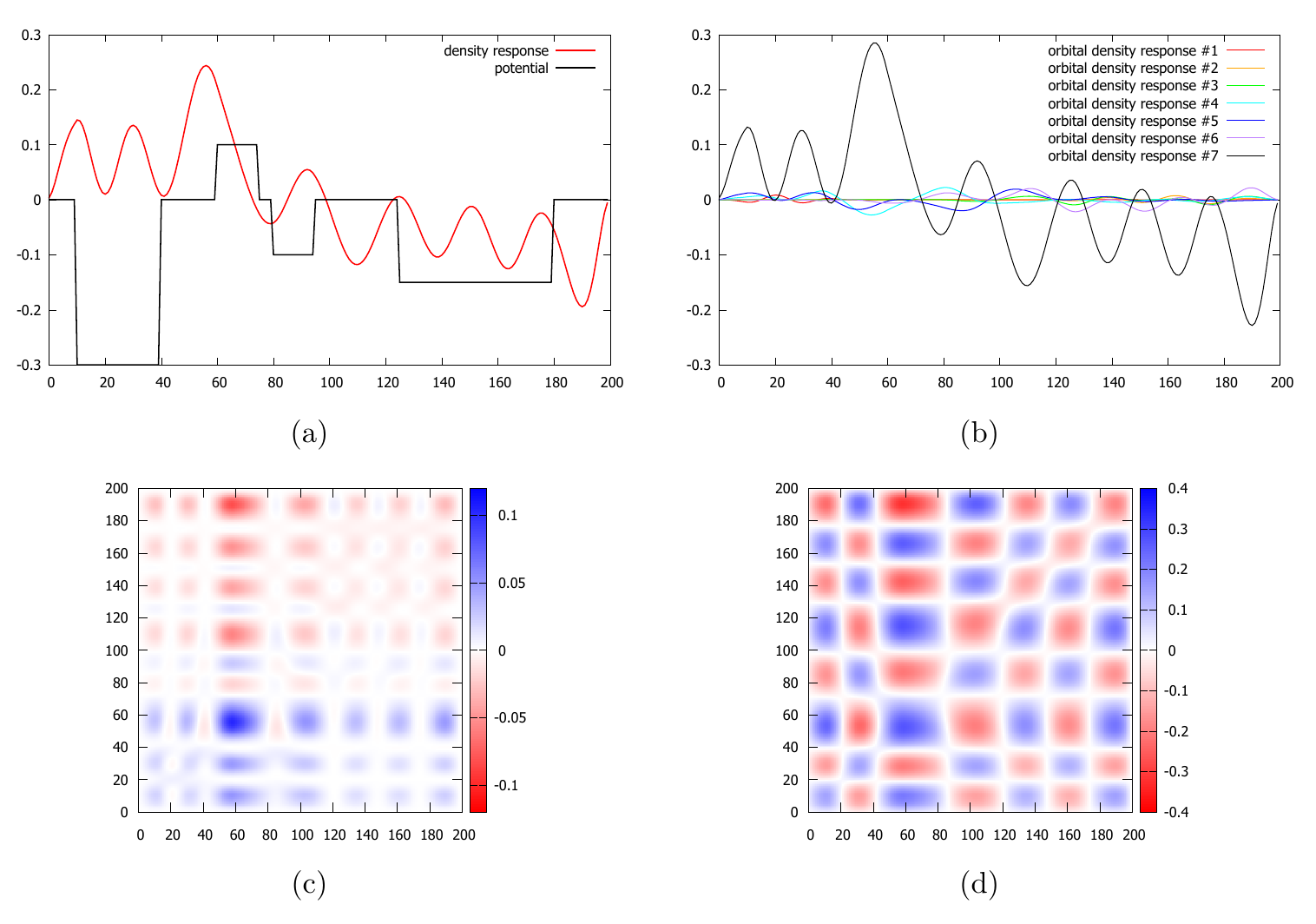}
\caption{(a) density response associated with the first excitation in a 1D system;
(b) individual density responses of the 7 occupied orbitals;
(c) associated PHM$_\omega$;
(d) associated TDM.} \label{fig:1DMol_ex1}
\end{figure}

Another simple 1D example is given in Fig. \ref{fig:1DMol_ex1}. The potential has several barriers and wells which are meant to represent a 1D molecule-like system with 7 doubly occupied Kohn-Sham orbitals.  We see that the first excitation is dominated by a single orbital density response with a strong peak  around 60. There are clear features in the PHM$_\omega$  around the 60th column and 60th row. By contrast, the TDM does not allow one to extract any useful information.

\begin{figure}[t]
\centering
\includegraphics[width=0.95\linewidth]{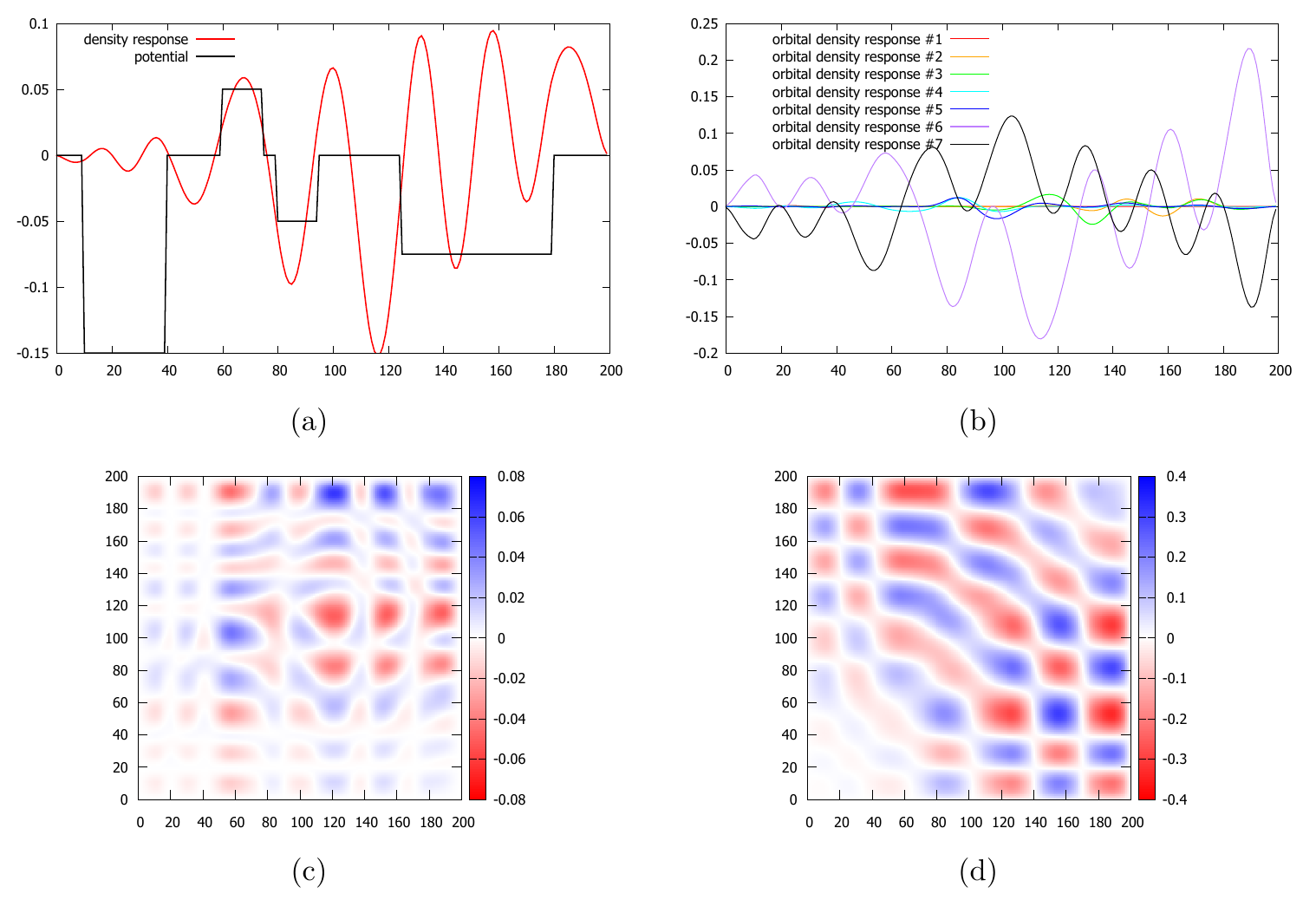}
\caption{As in Fig. \ref{fig:1DMol_ex1}, but for the second excitation.} \label{fig:1DMol_ex2}
\end{figure}

The second excitation, shown in Fig. \ref{fig:1DMol_ex2}, is selected because it is dominated by two orbital density responses, which makes it a more interesting case. We will use it as an example to compare the PHM$_\omega$ and  the PHM$_t$.

\begin{figure}[t]
\centering
\includegraphics[width=0.95\linewidth]{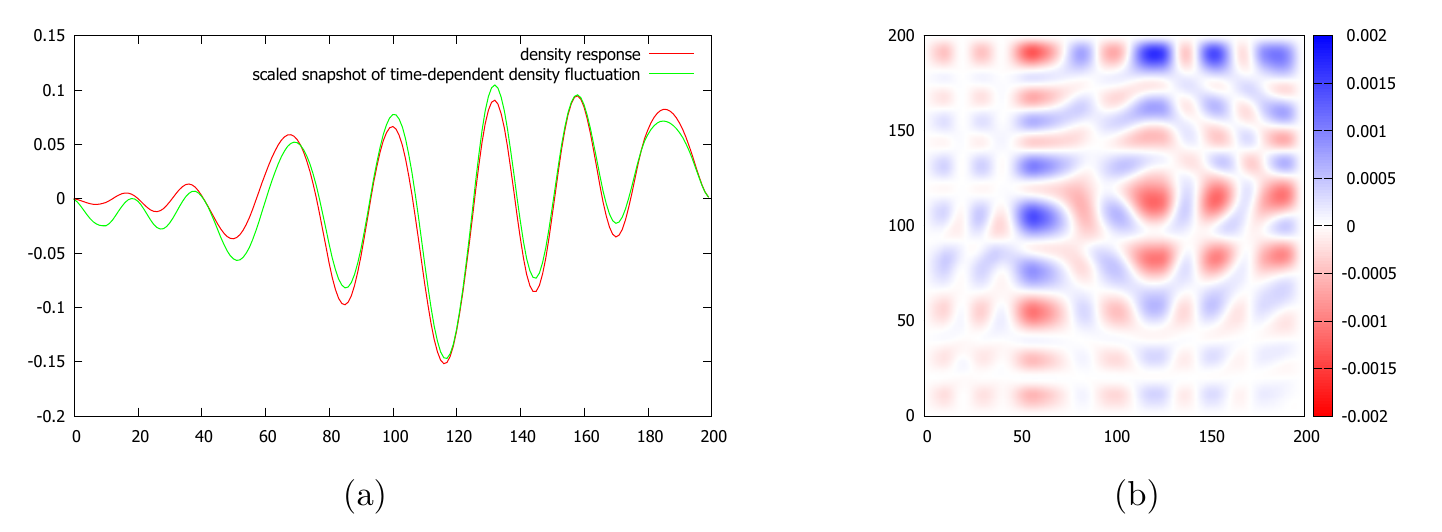}
\caption{(a) comparison of the total density fluctuation for the second excitation obtained from frequency-dependent linear response
and from time propagation (the snapshot is taken after the external field vanished);
(b) associated PHM$_t$. } \label{fig:1DMol_ex2_orbdensity}
\end{figure}

We solve the time-dependent Kohn-Sham equation with an external potential of the form
\begin{equation}
v_{\rm ext}(x,t)= \left\{ \begin{array}{cc}
V_{0}(x)A\sin^{2}\left(\frac{\pi}{w}t\right)\sin(\omega t), & 0<t<w, \\
0, & t>w .\end{array} \right.
\end{equation}
Here, $A\sin^{2}\left(\frac{\pi}{w}t\right)$ defines the time envelope of the field. The width $w$ of the envelope is selected to include 30 cycles of the oscillation, and the frequency $\omega$ is chosen as the second excitation energy. We found that $V_0(x)$, the spatial shape of the external field, only has
a very minor influence on the density response. The scaled time-dependent density fluctuation agrees with the linear response very well, as shown in the left panel of Fig. \ref{fig:1DMol_ex2_orbdensity}.
The right panel of Fig. \ref{fig:1DMol_ex2_orbdensity} shows the PHM$_t$. We observe a very nice agreement of the time-dependent snapshot
with the frequency-dependent map PHM$_\omega$ of Fig. \ref{fig:1DMol_ex2}.

\subsection{Representations for 3D systems}

For real systems such as molecules and solids, $\Xi$ is a complicated
function of six spatial coordinates. To extract any visual information from the PHM, simplification and
reduction to lower dimensionality is necessary. This can be done in various ways, depending on the system under
consideration and on the computational approach.

For instance, one  can plot $\Xi(\bfr_0,\bfr',t)$ or $\Xi(\bfr_0,\bfr',\Omega_n)$ as a
function of $\bfr'$ in a given plane cutting through a  molecule or  solid, where both $\bfr_0$ and $\bfr'$
are restricted to lie in the plane and $\bfr_0$ is a fixed reference point. This produces a direct 2D image of the PHM,
analogous to the way in which exciton wave functions and electron-hole distributions in periodic systems are typically
plotted \cite{Hummer2004,Varsano2008,Sharifzadeh2013}. A disadvantage of this way of representing the PHM is
that this limits the information one can extract to one origin, $\bfr_0$, at a time.

Alternatively, one can reduce the complexity of $\Xi$ by spatial averaging or coarse-graining procedures.
For instance, the Kohn-Sham molecular orbitals are often expressed in an atom-centered basis,
\begin{equation} \label{eq3.16}
\varphi_j(\bfr,t) = \sum_{l=1}^{N_A} \sum_{k=1}^{M_l} \zeta_{jk}^l(t) \eta_k^l(\bfr) \:,
\end{equation}
where $N_A$ is the number of atoms in the molecule and $M_l$ is the number of basis functions in the $l$th atom.
If this expansion is inserted into Eqs. (\ref{eq3.1}), (\ref{eq3.9}) or (\ref{eq3.10}),
a complication arises because the Kohn-Sham orbitals enter quadratically,
which leads to cross-terms between basis functions belonging to different atoms. To get rid of these terms,
one can make the approximation of zero overlap, $\eta_k^l(\bfr)\eta_{k'}^{l'}(\bfr)\approx 0$ for $l\ne l'$.
This allows one to represent the time- and frequency-dependent PHMs as $\Xi_{lm}(t)$ and $\Xi_{lm}(\Omega_n)$,
respectively, where the indices $l$ and $m$ run over all atomic origins and destinations of particles and holes.
The result is a two-dimensional array which is analogous to the TDMs found in the literature.
\cite{Mukamel1997,Tretiak1998,Tretiak2002,Tretiak2003,Tretiak2005,Wu2006,Igumenshchev2007,Li2007,Kilina2008,Wong2009,Xia2012}

Instead of using basis sets, some codes solve the Kohn-Sham equations on a real-space grid, in which case it is convenient
to work with a real-space partitioning scheme. We have implemented such a scheme for the {\tt octopus} code,\cite{octopus}
as we will now discuss.

Any spatial partitioning can be defined in terms of a set of weight functions, where different weights are assigned to different spatial
locations. Consider a function $f(\bfr)$ which is defined over all space;
we define a partitioning operator $\hat{P}_q$ acting on $f(\bfr)$ as
\begin{equation}\label{partition-general-define-1}
\hat{P}_q[f(\bfr)] =\int d^3r \: w_q(\bfr)f(\bfr) \:.
\end{equation}
We assume the weight function $w_q(\bfr)$ to be 1 if $\bfr$ lies within the $q$th bin; else, $w_q(\bfr)=0$. Hence,
$\hat P_q$ simply performs an average of $f(\bfr)$ over the $q$th bin. The
weight functions are assumed to seamlessly cover all space,
$\sum_{q=1}^{N_B}w_q(\bfr) =1$, where $N_B$ is the number of bins.

\begin{figure}[t]
\centering
\includegraphics[width=0.95\linewidth]{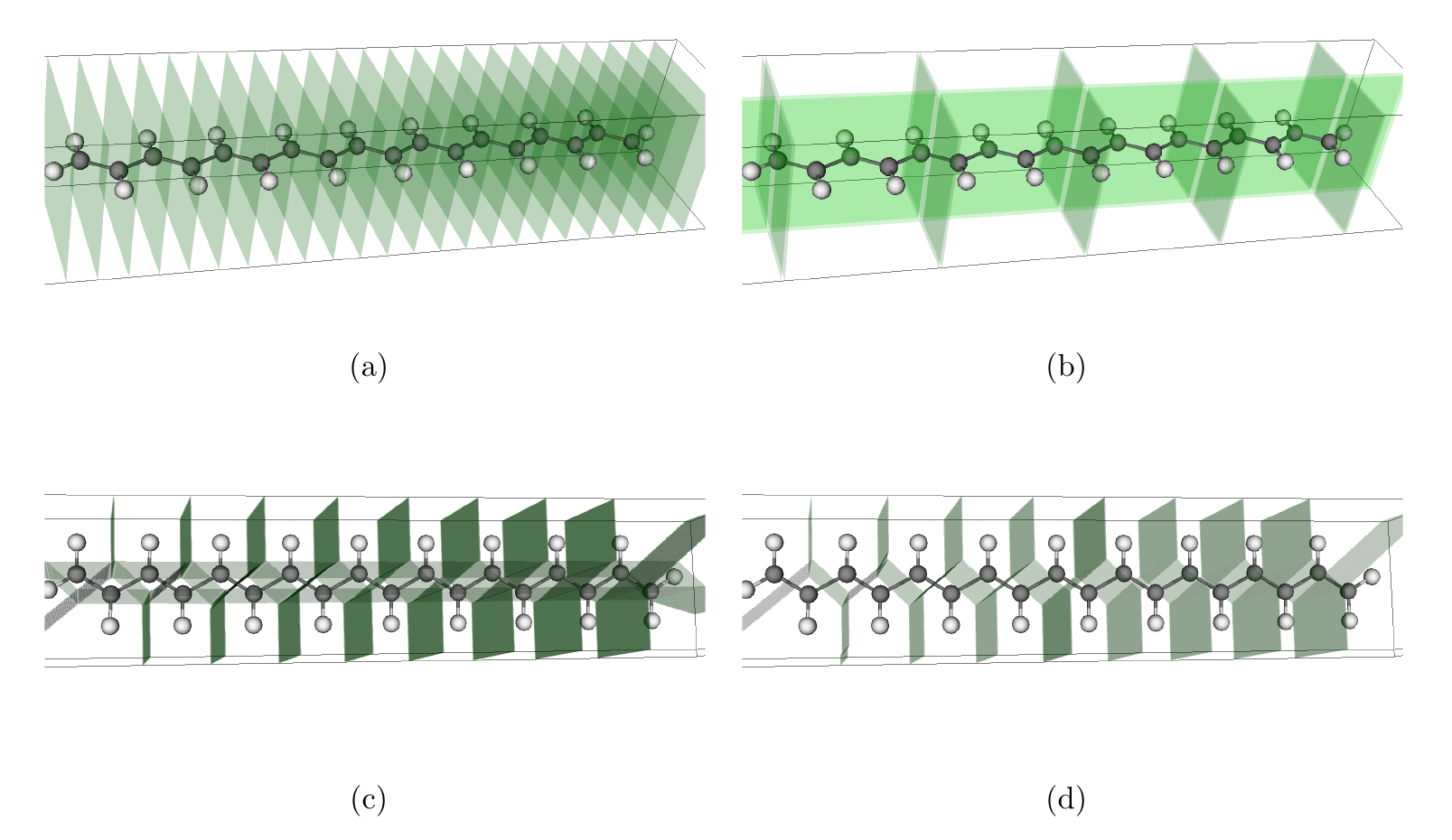}
\caption{Schematic illustrations of spatial partitioning schemes for molecules: (a)-(b) box partitioning,
 (c) atom centered partitioning. (d) atom centered partitioning with Hydrogen ignored.} \label{fig:partition}
\end{figure}

Figure \ref{fig:partition} shows examples of spatial partitioning for a linear molecule. The first two examples, (a) and (b), show a partitioning
into equally sized boxes, but with different degrees of coarse graining; this division of space is particularly simple, and yields a straightforward connection
between molecular geometry and spatial map. Clearly, this kind of box partitioning is most useful for molecules with
linear structure, and may become less practical for molecules of a more general (i.e., not linear) shape.

Having bins with identical sizes and shapes is conceptually straightforward but by no means necessary.
A spatial partitioning method with more flexibility, the atom centered partitioning, is shown in
Figs. \ref{fig:partition} (c) and (d). Here, we define the $q$th bin as that region in space
which is closer to the $q$th atom than to any other atom; this method is similar to the definition of the
Wigner-Seitz unit cell in a solid.

\subsection{The PHMs for a Hydrogen chain} \label{subsec:hydrogen_chain}

Let us consider a simple hydrogen chain with 8 H atoms. This is not a particularly interesting system, as far as the charge dynamics
during excitation processes is concerned, but it is simple and useful as a proof of concept to illustrate our approach.
We calculate the linear response and the time-dependent propagation using the adiabatic
local-density approximation (ALDA) for the xc potential and the xc kernel. The 3D objects are reduced by using the box partitioning scheme,
as discussed above, see Figs. \ref{fig:partition} (a) and (b).

\begin{figure}[t]
\centering
\includegraphics[width=0.95\linewidth]{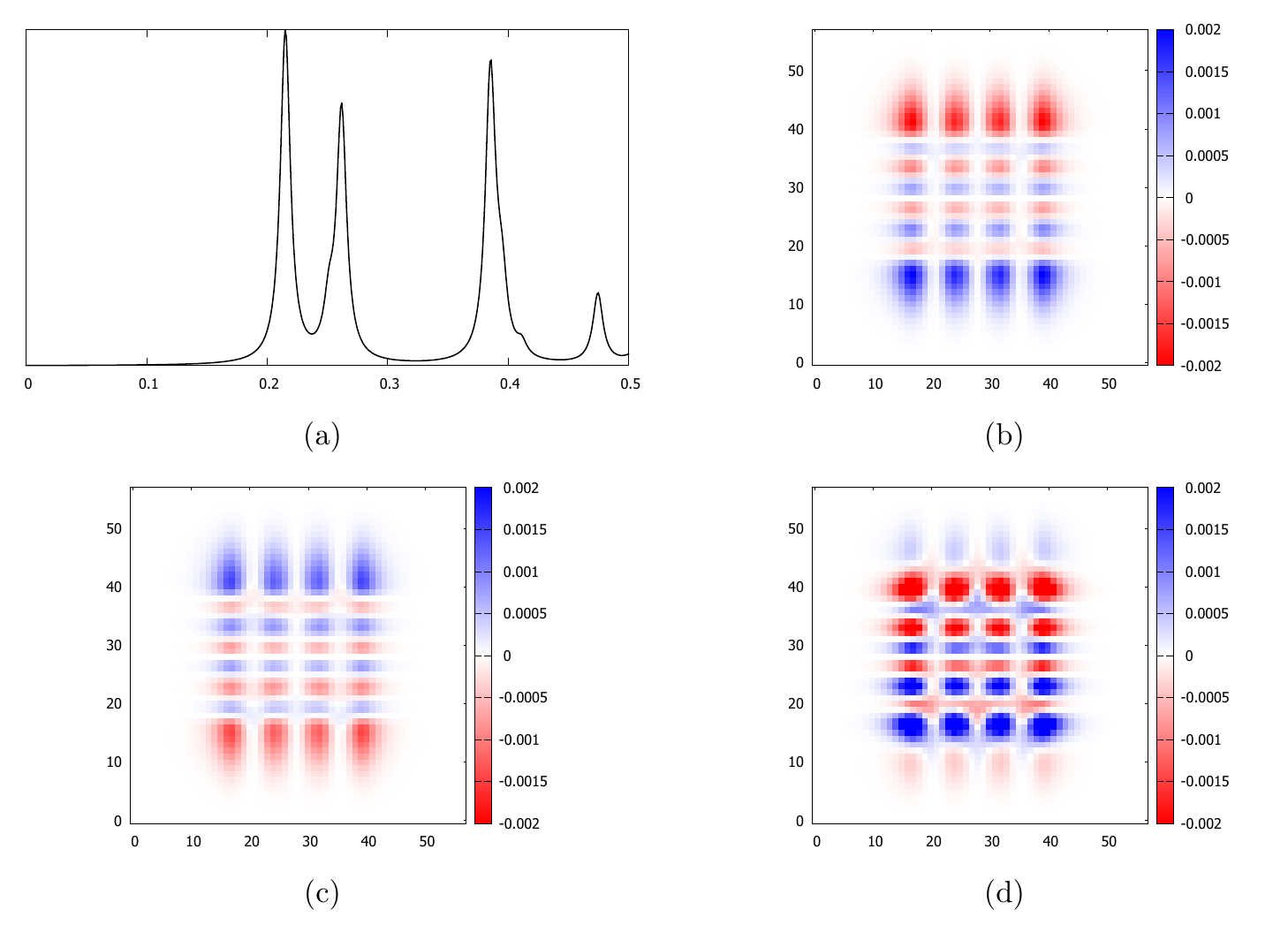}
\caption{(a) excitation spectrum of an H$_8$ linear chain, obtained using linear response with ALDA;
(b)-(d) PHM$_\omega$ for the first, second and third excitation peak.} \label{fig:3DH_freqPHM}
\end{figure}

In Fig. \ref{fig:3DH_freqPHM}, the excitation spectrum and the PHM$_\omega$ associated with the first three excitation frequencies are shown.
The first and the second excitation produce similar maps, with a significant accumulation of electron and hole
charges at the opposite ends of the chains. The third excitation is somewhat more confined to the interior of the chain.
Due to the delocalized nature of the molecular orbitals, the maps appear rather uniform, with no outstanding CT features
(as expected).

\begin{figure}[t]
\centering
\includegraphics[width=0.89\linewidth]{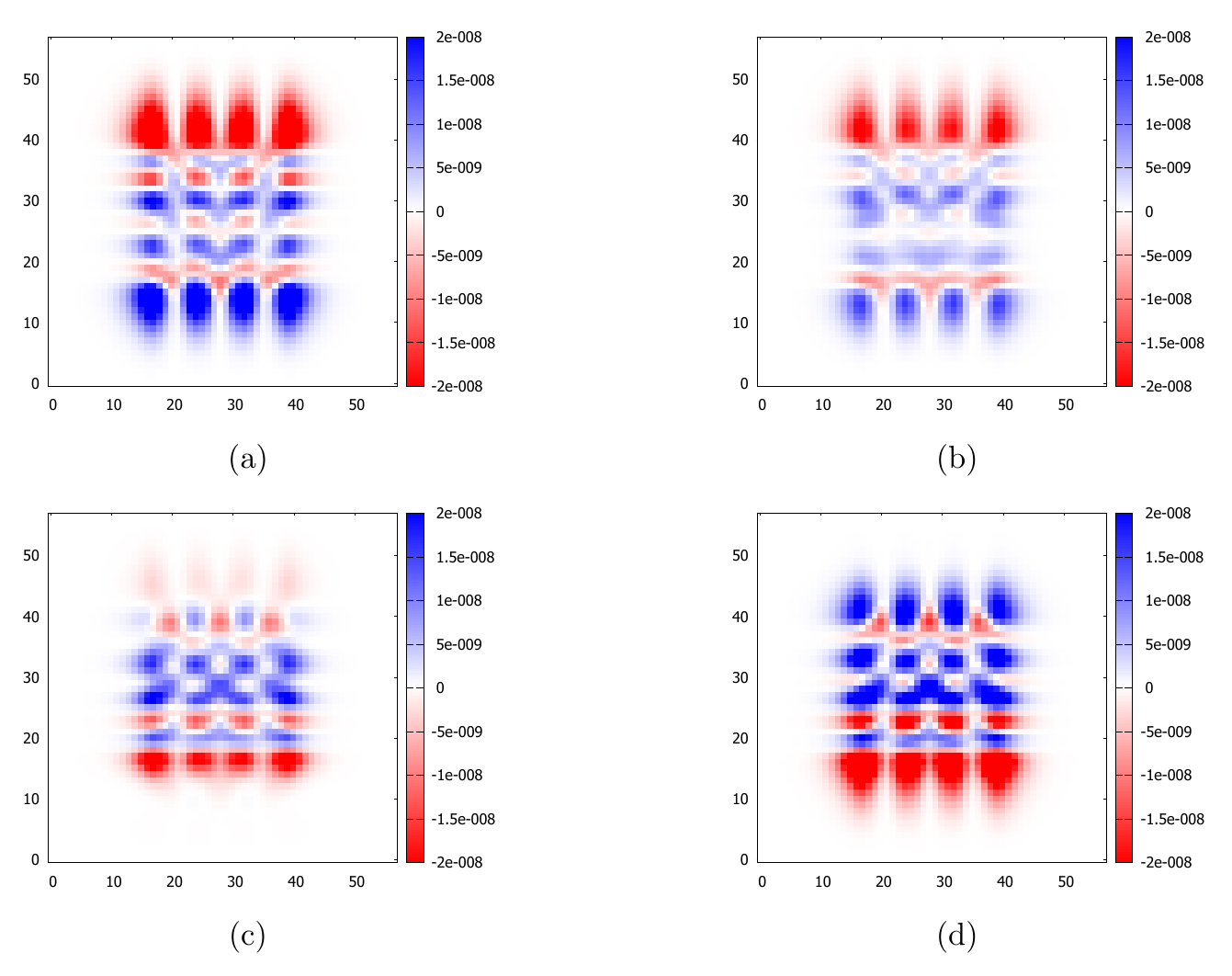}
\caption{A series of snapshots of the PHM$_t$ in an early stage of the simulation. The time-dependent
potential is in resonance with the first excitation of H$_8$. The snapshots start at 2.36 fs (a) and end at 2.47 fs (d)} \label{fig:3DH_tdPHM_series}
\end{figure}

Figure \ref{fig:3DH_tdPHM_series} shows four snapshots taken during the time propagation of the H$_8$ chain, subject to a laser pulse at
the first excitation frequency. The pattern is very similar to the corresponding PHM$_\omega$, Fig. \ref{fig:3DH_freqPHM} (b),
with a time periodicity associated with the density response. There is a periodic oscillation of charge density back and
forth along the chain, with a spatial modulation imprinted by the individual H atoms.

In this example, the driving external potential was chosen to be monochromatic and on resonance with a given excitation energy,
with a rather small amplitude and with a slowly changing pulse envelope,
in order to allow a comparison with the PHM$_\omega$ coming from frequency-dependent linear response theory.
However, in general no such restriction is necessary: the PHM$_t$ can be calculated for arbitrary time-dependent fields, as long as they are sufficiently weak so that the dynamics remains in the linear regime.

\subsection{The PHM$_\omega$ for a BT based molecule} \label{subsec:bt}

\begin{figure}[t]
\centering
\includegraphics[width=0.95\linewidth]{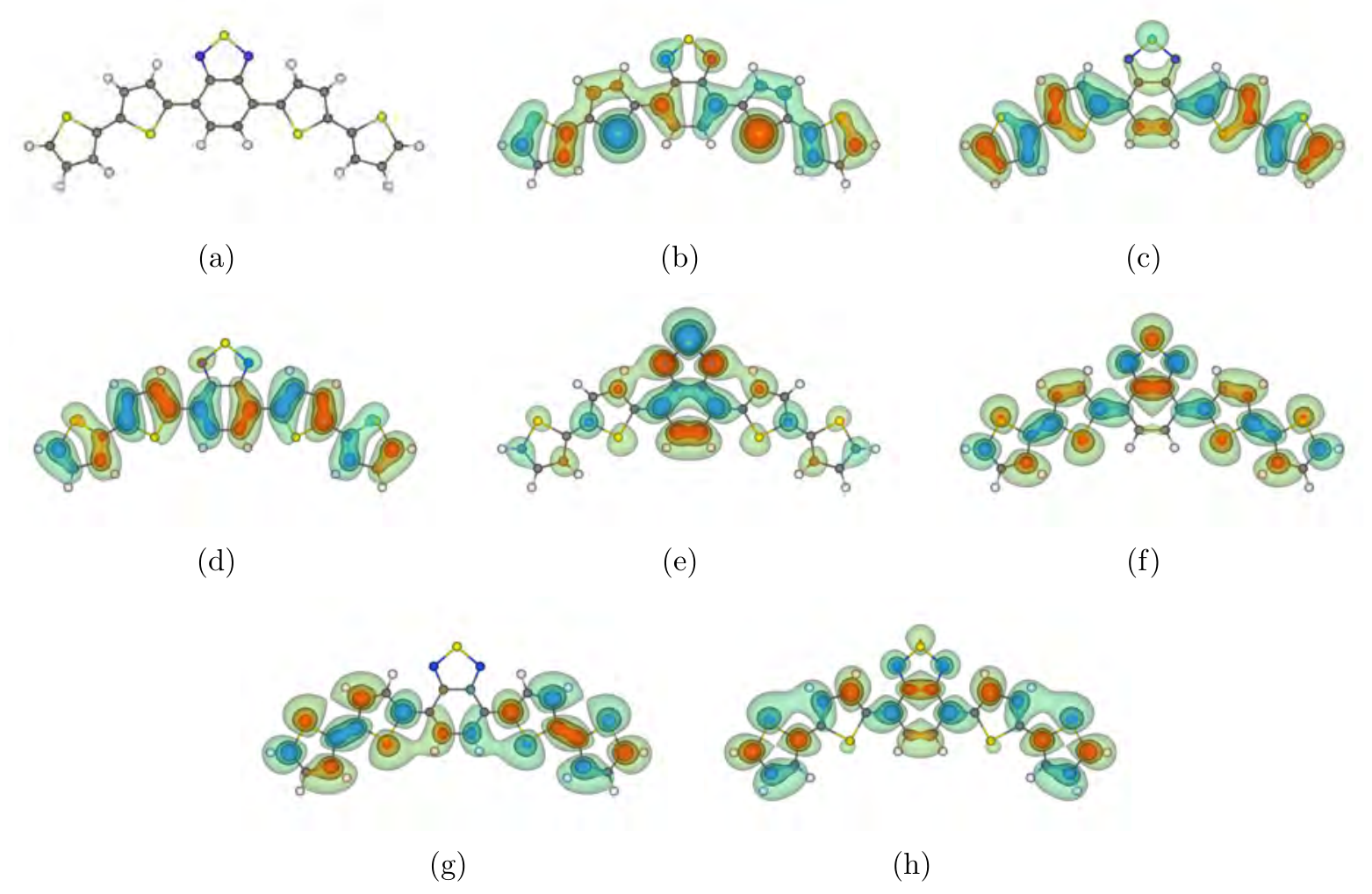}
\caption{(a) structure of Benzothiadiazole (BT); (b)-(h) HOMO-2 to LUMO+3.} \label{fig:BT_orbitals}
\end{figure}

We now apply the PHM$_\omega$ concept to a Benzothiadiazole (BT) based Donor-Acceptor-Donor (DAD) molecule, illustrated in Fig.  \ref{fig:BT_orbitals}.
This system belongs to a class of organic molecules which have been well studied for years, since they are known as promising materials for
photovoltaic applications\cite{Pop2015}. For our purposes, the BT DAD system is well suited because it will allow us to illustrate how
the PHM works for intramolecular CT processes.

Fig. \ref{fig:BT_orbitals} shows the structure of BT, including two thiophene units on each side (which are expected to act as donors); in practice,
systems of this kind may include side chains of various types, which we have here neglected. We also show the most important molecular
orbitals, from HOMO-2 to LUMO+3, calculated using the B3LYP hybrid functional.\cite{Stephens1994} The excited states were then calculated using ALDA.
Clearly, the orbitals extend over the entire system without concentration to any particular thiophene or the BT unit.

\begin{figure}[t]
\centering
\includegraphics[width=0.9\linewidth]{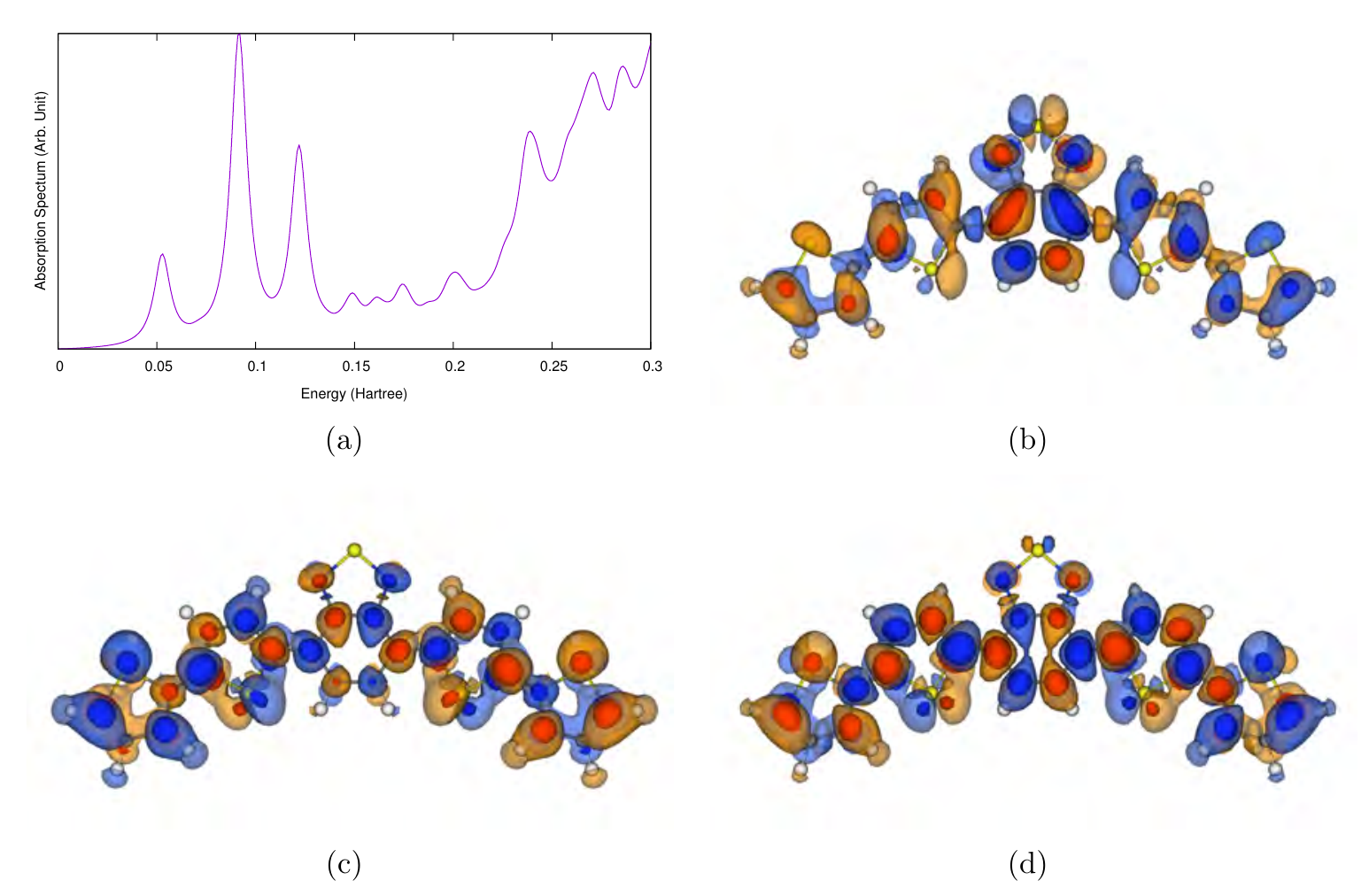}
\caption{(a) absorption spectrum of BT; (b)-(d) transition densties associated with the first three peaks located at 0.0527, 0.0915 and 0.122 Hartree
(red: density depletion, blue: density accumulation).} \label{fig:BT_casida}
\end{figure}

\begin{figure}[t]
\centering
\includegraphics[width=0.9\linewidth]{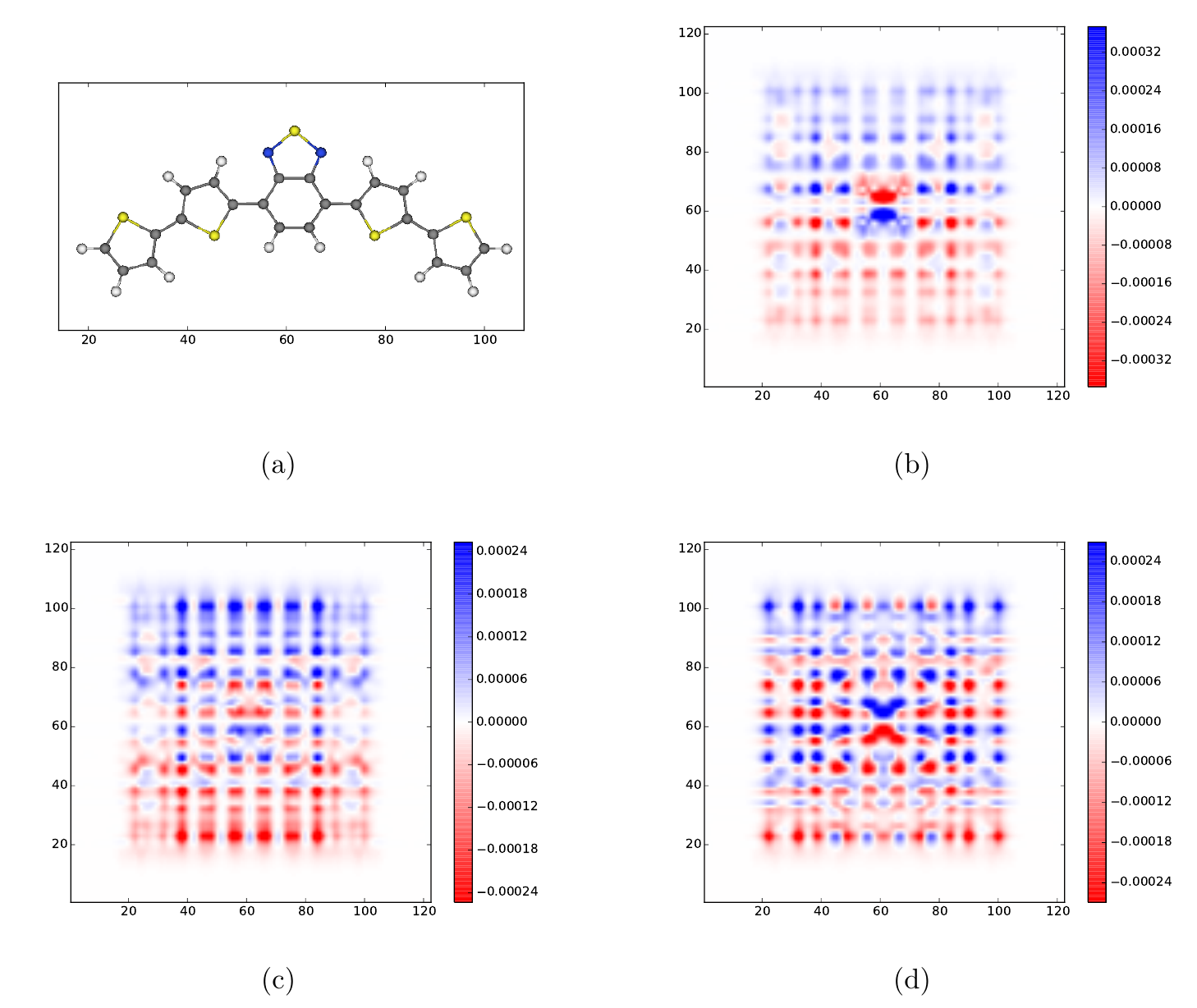}
\caption{(a) coordinates for the PHM box partitioning (Horizontal label defines the slicing along x direction); (b)-(d) PHM$_\omega$ for the first three excitations.} \label{fig:BT_PHM}
\end{figure}

The solution of the TDDFT linear
response equation (\ref{eq2.5}) yields the optical spectrum given in Fig. \ref{fig:BT_casida} (a). The figure shows that there
are three significant peaks at low energies; the associated transition densities are also given.
The transition densities hint at a pronounced polarization behavior along the entire system as well as between neighboring units;
let us now see whether the PHM$_\omega$ will give us more insight into the nature of the excitations.
The PHM$_\omega$s are shown in Fig. \ref{fig:BT_PHM}. They were again calculated using a simple partitioning scheme obtained by slicing along the
$x$-direction, see Figs. \ref{fig:partition} (a) and (b).

The PHM$_\omega$s for the first three excitations show intricate patterns of charge redistribution, and indicate that the BT central
unit and the thiophene side units play very different roles. For the lowest excitation, Fig. \ref{fig:BT_PHM} (a), we see that
there appears to be a rather uniformly polarized background, but the central part shows the reverse behavior.
The vertical columns around 20-30 and 90-100 suggest that the leftmost and rightmost thiophene units contribute uniformly
to the polarization of the system. However, the other two thiophene units (columns 40-50 and 75-85) are strongly
influenced by the vicinity to BT, and transfer a significant amount of electron and hole probability to the BT unit.

The PHM$_\omega$s for the second and third excitations can be analyzed along similar lines. In the second excitations,
the left and right thiophene units are not strongly involved and play a rather passive role; by contrast, the two central thiophenes and the BT
unit give rise to a significant electron-hole redistribution. The third excitation is even more complex, but again the BT unit
stands out.

Overall, the PHM$_\omega$s that are shown in Fig. \ref{fig:BT_PHM} do not exhibit a very strong intramolecular CT character;
this would, for instance, be the case if a vertical column starting at one end of the molecule would just end up at the middle,
rather than extend over the entire length of the molecule. A possible reason for the rather delocalized nature of the excitations
that we observe here is that the TDDFT calculations were done with the ALDA. Other, more sophisticated xc functionals (for instance, certain classes
of hybrid functionals) are known to better capture CT excitations. Further tests will be required to ascertain this.

\section{Conclusion}\label{sec5}

In this paper we have proposed a new computational tool, the Particle-Hole Map, which can be used to visualize not only the charge redistribution,
but the detailed origins and destinations of electrons and holes during an excitation process. It is hence ideally suited to identify and
characterize charge-transfer excitonic effects. The method is applicable to a wide range of materials, from molecules to solids.

We have defined the PHM both in the frequency and in the time domain, to be applied as a post-processing tool following a TDDFT calculation.
The PHM$_\omega$ and PHM$_t$ are defined in terms of Kohn-Sham orbital densities; this means that
they are not formally rigorous observables in terms of the density, but they rely on giving physical meaning to individual Kohn-Sham orbitals.
From a formal point of view this may be questioned, but there is a general consensus that
the Kohn-Sham orbitals are more than just auxiliary quantities and, indeed, have a considerable degree of physical significance,
\cite{Meer2014} in particular if they have been obtained with high-quality xc functionals.

The main appeal of the PHM$_\omega$ and PHM$_t$ is their interpretation in terms of joint probabilities of
ground-state orbital densities and excited-state density fluctuations. In view of this, we can simply say that
the PHM tells us where electrons and holes are coming from and where they are going to during an excitation process.
No other nonlocal visualization tool (such as the TDM) has such a simple interpretation. In addition, the PHMs satisfy physically important sum rules.

We have illustrated the PHMs with various examples. Simple 1D model systems teach us the rules how to read the PHMs, and show us
that the PHM$_\omega$ and PHM$_t$ are consistent with each other. In 3D systems, a dimensional reduction is necessary; we have
given several such schemes, depending on the particular mode of calculation. The majority of numerical calculations in molecules are carried
out with atom-centered basis sets, and it is straightforward to construct the PHMs in this case. Here, we carried out calculations using the
grid-based {\tt octopus} code, and for this purpose we introduced spatial partitioning schemes that amount to simple
binning. Applications to linear H$_8$ chains and to BT based donor-acceptor-donor systems illustrated the viability and
usefulness of the approach. More applications to other types of molecules and to periodic extended systems are currently in progress and will be reported elsewhere.

Here we have focused on applications of the PHM$_t$ in the linear regime, restricting the calculations
to weak and essentially monochromatic external fields in order to demonstrate that the time- and
frequency-dependent PHMs are consistent. However, the PHM$_t$ is suitable for arbitrary time-dependent
potential, as long as one remains in the linear regime; this opens up a wide field of application in the area of ultrafast dynamics.

To go beyond the linear regime, one could attempt to generalize the definition of the PHM$_t$ and include terms of order $\delta^2$.
However, in the nonlinear regime it is no longer justified to discard the contributions of the initially occupied orbitals
in the induced change $\delta \varphi_i(\bfr,t)$ of the $i$th Kohn-Sham orbital. In that case it would be preferable to work directly with
the joint probability $P(\bfr,\bfr',t)$ of Eq. (23).

Other future applications of the PHM may involve the motion of the nuclei, for instance using Ehrenfest dynamics.
Lattice relaxation plays an important role in many optical processes and is absolutely crucial for charge separation
in photovoltaics. \cite{Rozzi2013} It would be straightforward to implement the PHM$_t$ in systems with moving nuclei.
Extensions to magnetic systems, involving a spin-dependent PHM, are also possible.

\begin{acknowledgement}
This work was supported by National Science Foundation Grants No. DMR-1005651 and DMR-1408904.
We thank Suchi Guha for very valuable discussions about materials for organic photovoltaics.
\end{acknowledgement}


\end{document}